\documentstyle[12pt]{article}
\begin{document}
\newcommand{\goo}{\,\raisebox{-.5ex}{$\stackrel{>}{\scriptstyle\sim}$}\,}
\newcommand{\loo}{\,\raisebox{-.5ex}{$\stackrel{<}{\scriptstyle\sim}$}\,}
\vspace{0.2 truecm}
 
\begin{center}

{\Large \bf  The critical temperature of nuclear matter and 
fragment distributions in multifragmentation of finite nuclei.}
\end{center}

\begin{center}
{\large R.~Ogul$^{1,2}$ and A.S.~Botvina$^{1,3}$}
\end{center}

{\it
$^{1}$Gesellschaft fur Schwerionenforschung, D--64291 Darmstadt, Germany.

$^{2}$Department of Physics, University of Selcuk, 42079 Kampus, Konya, Turkey.

$^{3}$Institute for Nuclear Research, Russian Academy of Science,
117312 Moscow, Russia. 
}

\begin{abstract}
The fragment production in multifragmentation of finite nuclei is 
affected by the critical temperature of nuclear matter. 
We show that this temperature can be determined on the basis of 
the statistical multifragmentation 
model (SMM) by analyzing the evolution of fragment distributions 
with the excitation energy. This method can reveal a decrease of 
the critical temperature that, e.g., is expected for neutron-rich matter. 
The influence of isospin on fragment distributions is also discussed. 
\end{abstract}

\hspace{0.9 truecm}PACS numbers: 25.70.Pq, 21.65.+f
 
\vspace{0.5 truecm}

Properties of nuclear matter have been under investigation for several 
decades (see e.g. \cite{baym}). 
Besides their general interest for nuclear physics, 
these studies are very important for our understanding astrophysical 
objects, such as 
neutron stars. The information about nuclear matter in its ground state 
and at low temperatures is usually obtained as a theoretical extrapolation, 
based on nuclear models designed to describe the structure of real nuclei. 

%\vspace{0.5 truecm}

It is instructive to investigate the thermodynamical properties of 
neutron--rich 
matter under extreme conditions of low densities and high temperatures. 
This situation is expected, for example, at supernova II 
explosions and during the formation of neutron stars. We believe that 
the liquid-gas type phase transition is manifested, in this case, in the 
forms of instabilities leading to fragment production. 
The models used for extracting nuclear matter properties in the phase 
transition region should be capable of describing the disintegration of 
homogeneous matter into 
fragments. It is also important that they should allow to be tested in 
nuclear reactions leading to the 
total disintegration of real nuclei at high excitation 
energy. The multifragmentation reactions, which started to be investigated 
experimentally nearly 20 years ago (see reviews \cite{hufner85,bondorf95}), 
suit perfectly for this purpose. 

%\vspace{0.5 truecm}

Naturally, a thermodynamical model involved in this kind of analysis 
has to include the ingredients necessary for the description of nuclear 
matter and to 
provide a good reproduction of experimental data. So far, the statistical 
multifragmentation model (SMM) \cite{bondorf95} satisfies these 
requirements. The SMM has been designed to describe fragmentation and 
multifragmentation of excited finite nuclei \cite{bondorf85,botvina85}. 
It includes a liquid-drop approximation for individual fragments which 
corresponds to the liquid-drop description of nuclear matter 
\cite{baym,ravenhall83}. This is an essential difference of the SMM from 
other multifragmentation models, e.g. \cite{gross,randrup}, 
which do not take the nuclear matter properties explicitly into account. 
Examples of very successful applications of the SMM for the description 
of different experimental data can be found in 
[3,9--17]. 
%\cite{bondorf95,botvina95,dagostino96,williams,avdeev98,trautmann,wang99,dagostino99,scharenberg,srivastava}. 
Furthermore, the descriptions of data with 
statistical models confirm that multifragmentation of nuclei, despite of 
being a very fast process, proceeds under a high degree of thermalization. 

%\vspace{0.5 truecm}

Details of the SMM can be found in \cite{bondorf95}, here we concentrate on 
parts of the model which are important for the following discussion. 
The model describes the fragment formation at a low-density freeze-out 
($\rho \loo 1/3 \rho_0$, $\rho_0 \approx 0.15$fm$^{-3}$ is the normal 
nuclear density), where the nuclear liquid and gas phases coexist. 
The SMM phase diagram has already been 
under intensive investigations (see e.g. Ref.~\cite{bugaev}). 
The liquid-drop approximation suggests that the fragmentation process 
is accompanied by an increase of the surface of nuclear drops. The surface 
entropy contributes essentially to the statistical partition sum. We should 
point out that the surface free energy depends on the ratio of the 
temperature $T$ to the critical temperature of nuclear matter $T_c$. 
In the SMM the surface tension $\sigma (T)$ is given by 
\begin{equation}
\sigma (T) = \sigma (0) \Biggl(\frac{T_c^2-T^2}{T_c^2+T^2}\Biggr)^{5/4}. 
\end{equation}
This formula is obtained as a parameterization of the calculations of 
thermodynamical properties of the interface between two 
phases (liquid and gas) of nearly symmetric nuclear matter, which were 
performed with the Thomas--Fermi and Hartree--Fock methods 
by using the Skyrme forces \cite{ravenhall83}. In addition 
the scaling properties in the vicinity of the critical point 
(see \cite{landau}) were taken into account. 
At the critical temperature $T_c$ for the liquid-gas phase transition, 
the isotherm in the phase diagram has an inflection point. The surface 
tension vanishes at $T_c$ and only the gas phase is possible above 
this temperature. 
We emphasize that our analysis is based on this general effect and that 
our conclusions will remain qualitatively true in the case of other 
parametrizations satisfying this condition. 
This surface effect provides an effective way to study the influence of $T_c$ 
on fragment production in the multi-fragment decay of hot finite nuclei. 

%\vspace{0.5 truecm}

As was established by numerous studies (see e.g. [3--5,9,15,16]) 
%\cite{bondorf95,bondorf85,botvina85,botvina95,dagostino99,scharenberg}) 
the mass (charge) distribution of fragments 
produced in the disintegration of nuclei evolves with the excitation 
energy. At low temperatures ($T\loo 5$ MeV), 
there is a so-called $U$-shape distribution corresponding to partitions 
with few small fragments and one big residual fragment. 
This distribution looks like a result of an evaporative emission. At high 
temperatures ($T\goo 6$ MeV) the big fragments disappear, and there is an 
exponential-like fall of the mass distribution with mass number $A$. 
In the transition region $T\approx 5-6$ MeV, however, there is a smooth 
transformation of the first distribution into the second one. 
The mass distribution of intermediate mass fragments (IMFs, fragments 
with $A=5-40$) can be approximated by a power law 
$A^{-\tau}$ \cite{hufner85,bondorf95,goodman}. 
The $\tau$ parameter decreases with the temperature, goes through the 
minimum at $T\approx 5-6$ MeV, and then increases again. The small values 
of $\tau$ indicate that the probability for survival of the biggest 
fragment decreases drastically with the temperature. This behavior may be 
associated with a phase transition in finite systems. It has been shown 
in many studies (see e.g. 
\cite{bondorf95,dagostino99,srivastava} and references therein), that 
there are numerous peculiarities in this region, such as a plateau-like 
behavior of the caloric curve, large fluctuations of the temperature 
and of the number of the produced fragments, scaling laws for fragment 
yields, and other phenomena expected for critical behavior. 
Therefore, the temperature characterizing these phenomena is sometimes 
called the critical temperature for finite systems,  
and $\tau$ is considered as one of the critical exponents. 
The SMM can describe the critical behavior observed in the experiments 
[15--17]. However, in the present work we use the $\tau$ parametrization 
only for the characterization of shapes of the fragment mass distributions. 
In order to avoid any confusion with the 
standard definition of the critical temperature for nuclear matter, 
we note, in the following, the temperature corresponding to the 
critical phenomena as a break-up temperature for the disintegration of 
finite nuclei \cite{botvina85}. 

%\vspace{0.5 truecm}

The decrease of the surface energy with increasing $T/T_c$ (see formula (1)) 
influences the fragment production and, therefore, can be observed in the 
fragment distributions. In this paper we show that this effect can be 
used for the evaluation of $T_c$ by finding the minimum $\tau$ parameter 
$\tau_{min}$ and the corresponding temperature $T_{min}$. 
The physics behind the phenomenon is quite transparent: If the 
contribution of the surface energy is rapidly decreasing, a nucleus prefers 
to disintegrate into small fragments already at low temperatures. 
Simultaneously, fluctuations of size of the fragments increase 
considerably. As a result the mass distribution becomes flatter in the 
transition region, and this leads to a decrease of $\tau$.

%\vspace{0.5 truecm}

The SMM calculations were carried out for the $Au$ nucleus ($A_0$=197, 
$Z_0$=79) at different excitation energies and at a freeze-out 
density of one-third of the normal nuclear density. This 
choice is justified by the previous descriptions of the experimental data 
obtained for peripheral collisions [9,12--16]. 
%\cite{botvina95,avdeev98,trautmann,wang99,dagostino99,scharenberg}.
Below we present results as a function of both the temperature 
and the excitation energy, since they are related quantities 
\cite{bondorf95}. 

%\vspace{0.5 truecm}

We have started by using the standard value of the critical temperature 
implemented in the SMM, $T_c$=18 MeV. This value is consistent with many 
theoretical studies \cite{ravenhall83,ogul}. 
In Fig.~1 we show typical mass and IMF charge distributions, 
$\langle N_A \rangle \sim A^{-\tau}$ and 
$\langle N_Z \rangle \sim Z^{-\tau_z}$, 
at an excitation energy $E_{x}$=7 MeV/nucleon. One can see 
from this figure that the extracted $\tau$ and $\tau_z$ values are very 
close to each other, since the neutron-to-proton 
ratio of produced IMFs changes very little within their narrow charge 
range \cite{botvina95,botvina01}. 
As seen from Fig.~2, the dependences of these parameters versus excitation 
energy are nearly the same. The parameters obtained for primary hot 
fragments (excited nuclear matter drops) and after their secondary 
de-excitation (measured cold fragments) are also shown in this figure. 
One can see that the difference between the two cases is smallest around 
the minimum $\tau$ parameter. Therefore, the 
$\tau_{min}$ point is weakly affected by secondary processes. 

%\vspace{0.5 truecm}

The critical temperature reflects the properties of nuclear matter, however, 
these properties depend on the composition of this matter. For example, 
$T_c$ tends to decrease for neutron rich matter \cite{muller}. As it was 
discussed by many authors, see e.g. \cite{bonche}, the critical temperature 
can be traced back towards neutron rich matter by studies of the disassembly 
of nuclei far from stability. 
We consider $T_c$ as a free parameter in the SMM and analyze how $\tau$ and 
the break-up temperature can change. In Fig.~3 we show the results 
for $T_c=$10 and 30 MeV. In these cases 
the evolution of $\tau$ with the excitation energy is similar to the 
one shown in Fig.~2. 
However, the values of $T_{min}$ and $\tau_{min}$ are essentially 
different. This reflects a considerable change of masses for the 
dominating fragments. 
These values are plotted in Fig.~4 versus the critical temperature. 
It is seen that both parameters increase with $T_c$ and that they 
tend to saturate at $T_c \rightarrow \infty$ corresponding to the case of the 
temperature-independent surface. This behavior is expected, since in this case 
only the translational and bulk entropies of fragments, but not the surface 
entropy, influence the probability for the fragment formation. 

%\vspace{0.5 truecm}

In the case of neutron-rich matter, the contribution of the symmetry energy 
increases considerably. 
It is necessary to take into account the standard dependence of the $\tau$ 
parameter on the isospin of the source, while searching for $T_c$. 
We performed SMM calculations of 
multifragmentation of $^{124}Sn$ and $^{124}La$ nuclei, which 
can be used in experiments \cite{kezzar}, and compare them 
with the results obtained for $Au$ nuclei. This $Sn$ nucleus is nearly as 
neutron-rich as the gold nucleus, while the $La$ nucleus is 
neutron-poor. We have used 
the same model parameters as for the $Au$ case, with the standard $T_c=$18 
MeV. It is seen from Fig.~5 that the neutron-poor source results in 
slightly lower microcanonical temperatures in the transition region 
($E_x \approx 3-5$ MeV/nucleon). In Fig.~5 we show also 
the evolution of the $\tau$ parameter with the excitation energy. 
The results for $Au$ and $Sn$ are very 
similar and different from those obtained for the $La$. 
One can conclude, that 
IMF distributions approximately scale with the size of the sources, 
and that they depend on the neutron-to-proton (N/Z) ratios of the sources. 
This is because the symmetry energy still dominates over the Coulomb 
interaction energy for these intermediate-size sources. 
One can see from Fig.~5 that the source with the lower N/Z ratio 
leads to smaller $\tau$ parameters, i.e. to the flatter fragment 
distribution. We can explain this as an effect of the isospin 
(i.e. the symmetry energy) on fragment formation: A high N/Z ratio of 
the source favors the production of big clusters, since they have 
a large isospin. Therefore, in the transition region, 
partitions consisting of small IMFs and a big cluster dominate. 
This leads to a very prominent U-shape distribution with large $\tau$. 
When the N/Z ratio is low, the probability for a big cluster to survive 
is small and the system can disintegrate into IMFs, which have a 
favorable isospin in this case. The dominant fragment partitions tend to 
include IMFs of different sizes, and the fragment distribution is 
characterized by small $\tau$. Since the 
difference in the $\tau$ parameters between the neutron-rich and 
neutron-poor sources is quite large, it can be easily identified. 
The evolutions of fragment mass distributions caused by decreasing 
$T_c$ and changing N/Z ratio can interfere and, therefore, the influence 
of the critical temperature can be separated only after the comparison of 
experimental data with calculations.

%\vspace{0.5 truecm}

In view of these theoretical findings it is instructive to 
demonstrate the possibility of the application of such an approach for the 
analysis of experimental data. Presently, there are several experimental 
analyses aimed at the extraction of the critical exponent $\tau$ in reactions 
with $Au$ nuclei \cite{dagostino99,srivastava,elliott}. Those methods are 
not equivalent to the one suggested above, however, they are related 
to the fragment distributions, and the critical exponents can be used 
for an estimation of $\tau_{min}$ and $T_c$. 
The extracted break-up excitation energies $E_x$ vary within 3.8 to 4.5 
MeV/nucleon, while $\tau$ is in the range from 2.12 to 2.18. 
As seen from Figs.~2 and 3, for $T_c \geq$ 18, at these excitation energies 
$\tau$ is larger than $\tau_{min}$. 
We have performed an interpolation of the SMM calculations for the Au sources 
and found that they fit the experimental values of $E_x$ and $\tau$ if 
$T_c$ is in the range between 18 and 22 MeV. We have also seen from our 
analysis that $\tau_{min}$ is lower than the extracted $\tau$ by around 15\%. 
The $T_c$ estimated in this way is very close to the standard SMM 
parametrization. 
This conclusion is supported by the analyses of Refs.~[15--17] 
%Refs.~\cite{dagostino99,scharenberg,srivastava} 
showing that the standard SMM reproduces 
both the experimental critical exponents and other characteristics of 
produced fragments.  
It is interesting that in Ref.~\cite{srivastava} a small critical exponent 
$\tau \approx$ 1.88 is reported for multifragmentation of $Kr$ 
nuclei, which have N/Z ratios lower than $Au$ nuclei. This is 
an indication of the importance of the isospin effects, as discussed above. 
It is worth noting that 
recently a very close critical temperature ($T_c \approx 16.6$ MeV) was 
extracted from analysis of the break-up ("limiting") temperatures in 
Ref.~\cite{natowitz}. It is also in agreement with the temperature 
$T_c \approx 20$ MeV obtained in the experiment of Ref.~\cite{karnaukhov}. 
However, it would be important to identify regular changes of 
$T_c$, which needs involving new sources with different isospins. 

%\vspace{0.5 truecm}

In summary, we have pointed out that within the SMM the critical temperature 
of nuclear matter can influence the fragment production in 
multifragmentation of nuclei through the surface energy. We have suggested 
that this influence can be observed in the $A^{-\tau}$ parameterization of 
the fragment yields by finding the minimum $\tau$ parameter. 
In the experiments the measured values of the parameters are 
consistent with the standard SMM assumption and slightly 
higher values of the critical temperature, $T_c \approx 18-22$ MeV. 
The SMM predicts that variations of 
$\tau_{min}$ are especially large in the region of low $T_c$. 
Therefore, there is a possibility to investigate the decrease of $T_c$ for 
nuclear matter under extreme conditions, by studying the evolution of 
$\tau_{min}$ and $T_{min}$ in the multifragmentation of finite nuclei. 
This could be realized in the case of the neutron-rich nuclei delivered 
by current accelerators with radioactive heavy-ion beams \cite{kezzar}. 
The isospin of the source influences also the 
fragment production through the symmetry energy. 
We have demonstrated how the isospin affects the fragment distributions, 
that should be taken into account in these studies. 

%\vspace{0.5 truecm}

The authors thank GSI for hospitality and support. We appreciate 
stimulating discussions with V.A.Karnaukhov and I.N.Mishustin.  
We are very indebted to W.~Trautmann and J.~Lukasik for discussions 
and help in preparation of the manuscript. 
R.~Ogul acknowledges financial support of TUBITAK-DFG cooperation.

\vspace{0.5 truecm}

{\large \bf  {Figure captions}}\\

{\bf Fig.1:} {Average fragment mass and charge yields 
$\langle N_A \rangle$ and $\langle N_Z \rangle$, 
after multifragmentation of $Au$ nuclei at 
an excitation energy of 7 MeV/nucleon. Solids lines are 
$\sim A^{-\tau}$ and $\sim Z^{-\tau_z}$ fits of the IMF yields. 
}
\vspace{0.5 cm}

{\bf Fig.2:} {Evolution of $\tau_z$ (top panel) and $\tau$ 
(bottom panel) parameters with the 
excitation energy $E_x$ of $Au$ sources calculated with the 
standard SMM parameterization. The open circles are for hot primary 
fragments and the full squares are for observed cold fragments.
}
\vspace{0.5 cm}

{\bf Fig.3:} {Evolution of $\tau$ parameters in the SMM 
for cold fragments with 
the excitation energy $E_x$. The top panel is for the critical 
temperature $T_c$=10 MeV, the bottom panel for $T_c$=30 MeV. 
}
\vspace{0.5 truecm}

{\bf Fig.4:} {The minimum $\tau_{min}$ parameter for cold fragments 
(top panel) and the corresponding temperature $T_{min}$ (bottom panel) 
as function of the critical temperature $T_c$ of nuclear matter.
}
\vspace{0.5 truecm}

{\bf Fig.5:} { 
The temperature (top panel) and $\tau$ parameters for cold fragments 
(bottom panel) versus excitation energy in multifragmentation. 
The solid, dashed and dotted lines are SMM calculations 
performed for sources with different sizes or isospin (see the figure) 
with the standard $T_c$=18 MeV. 
}

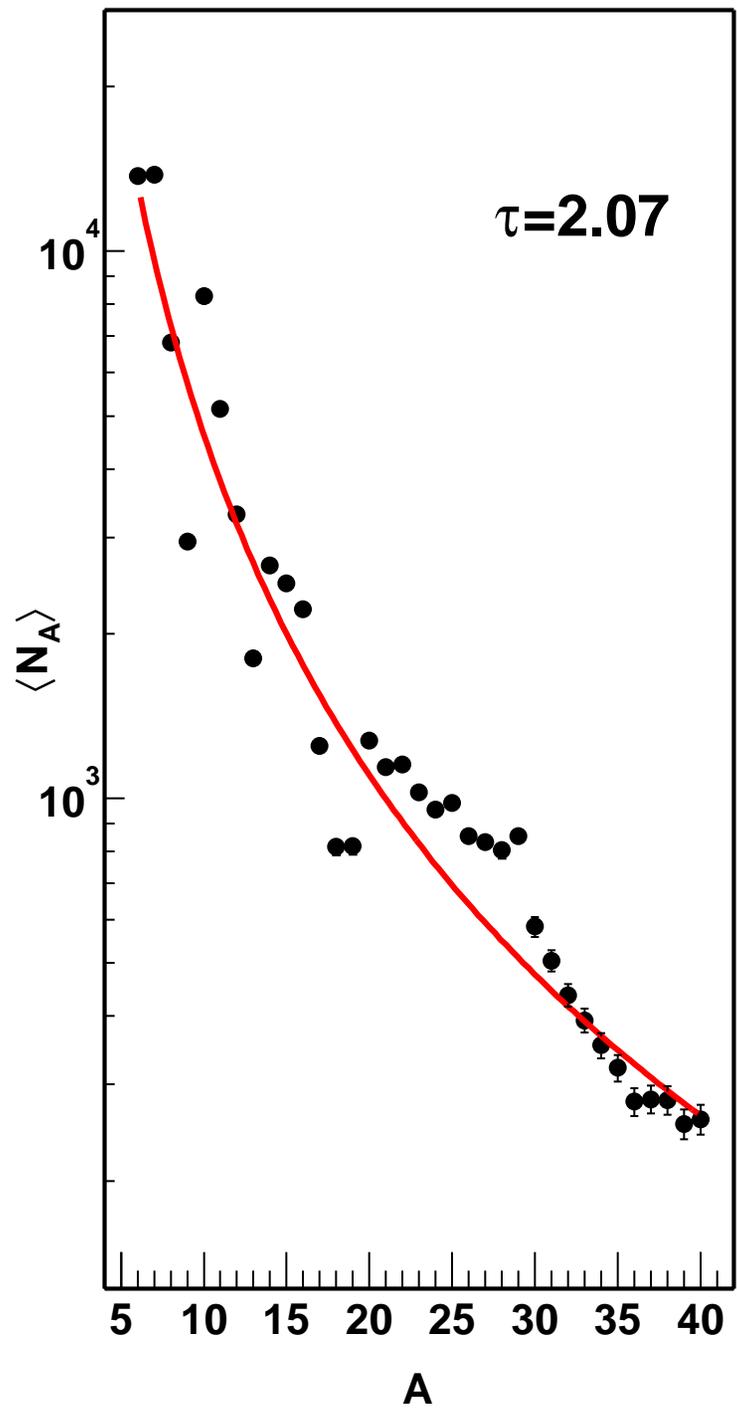
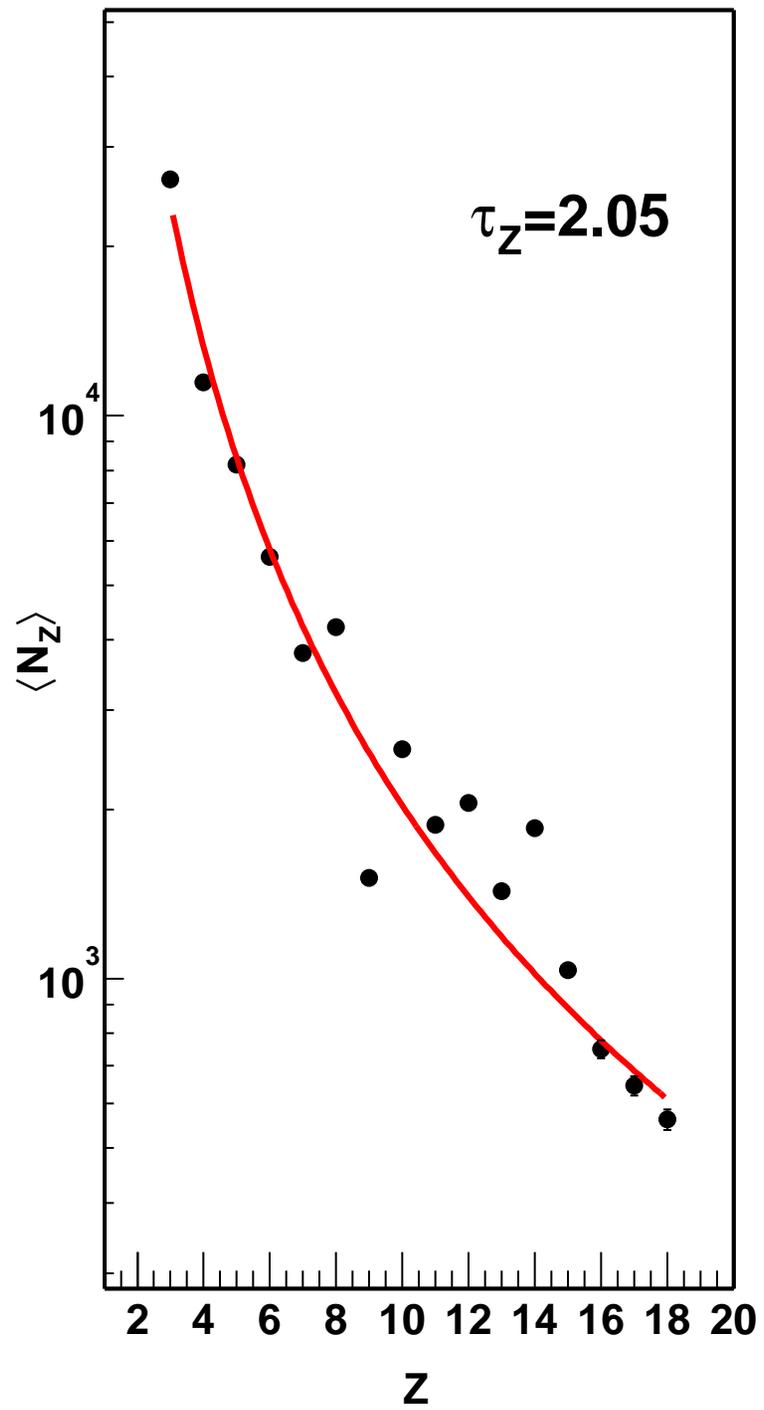
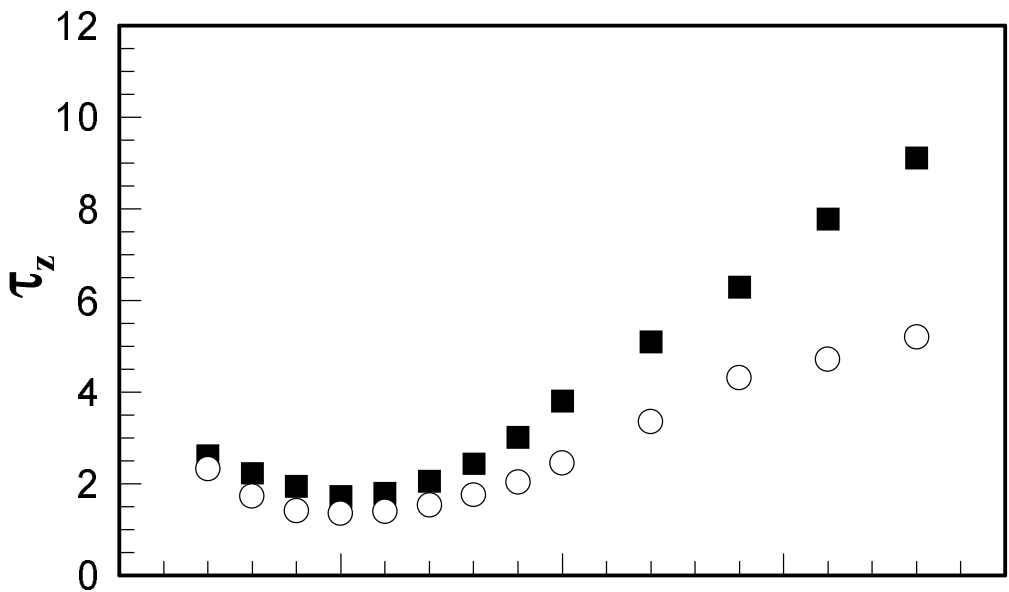
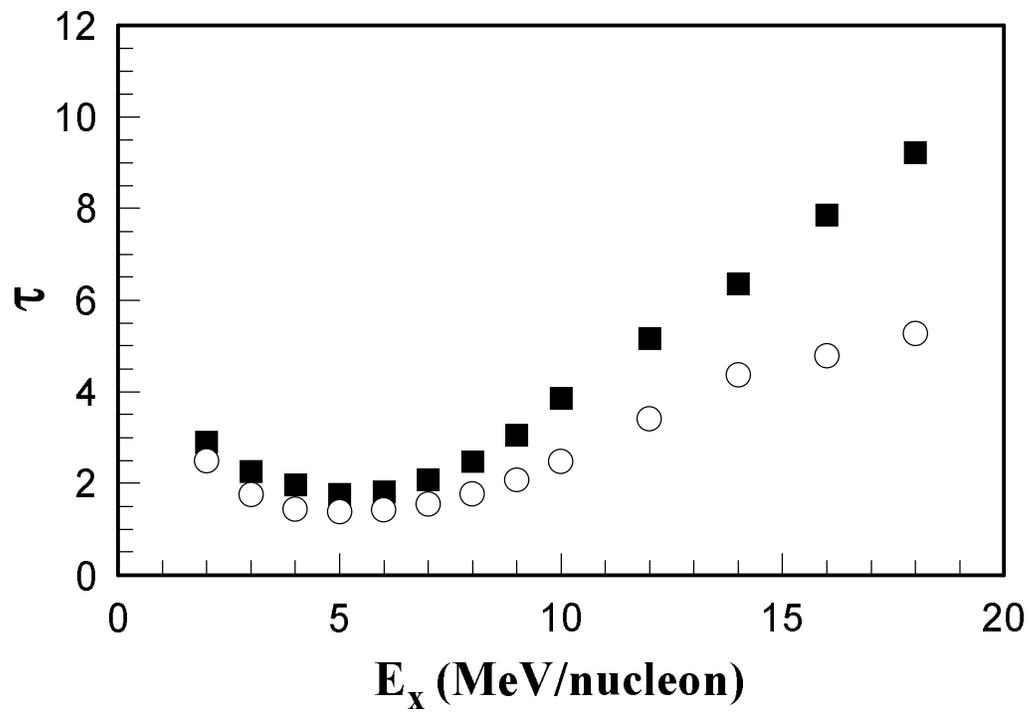
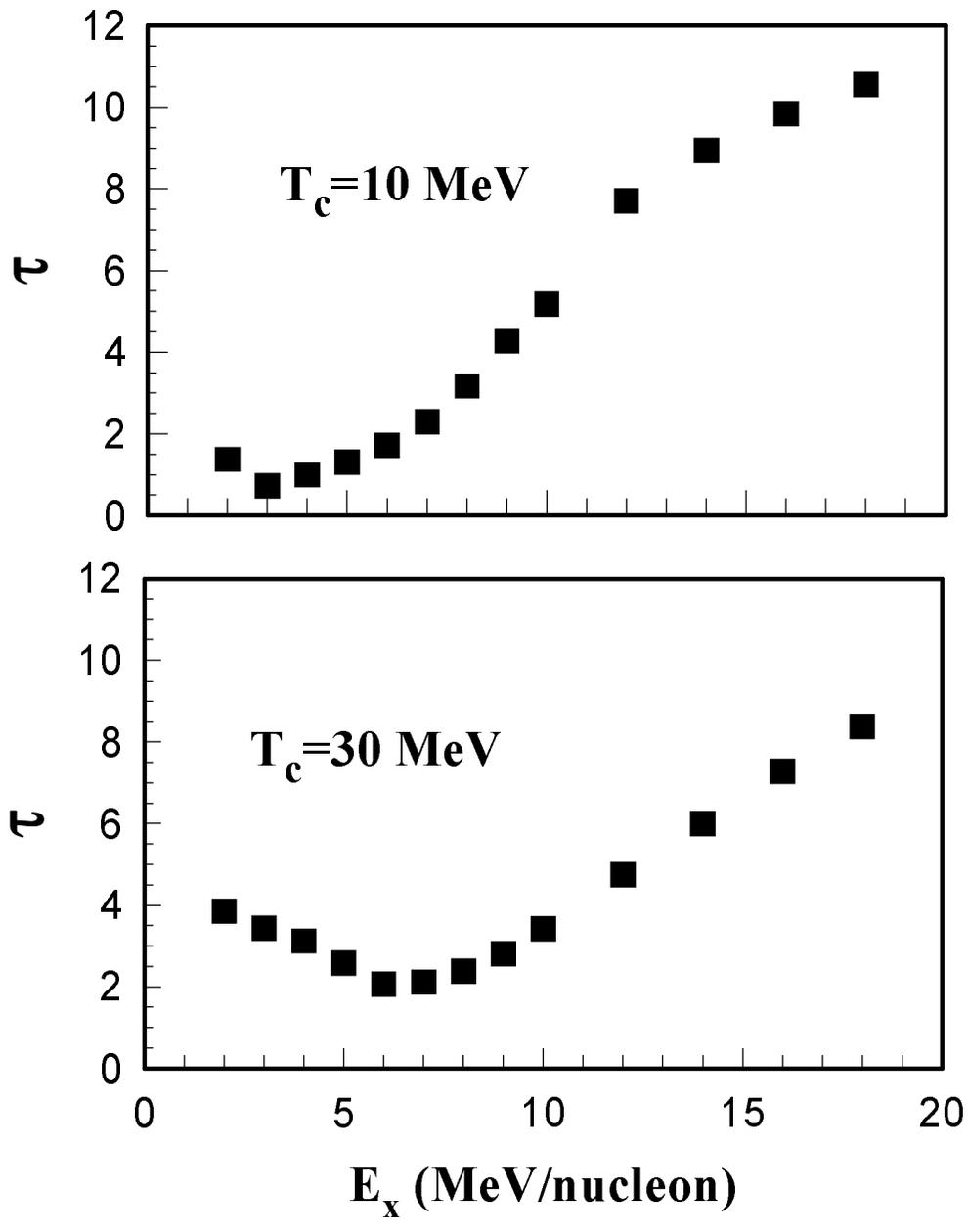
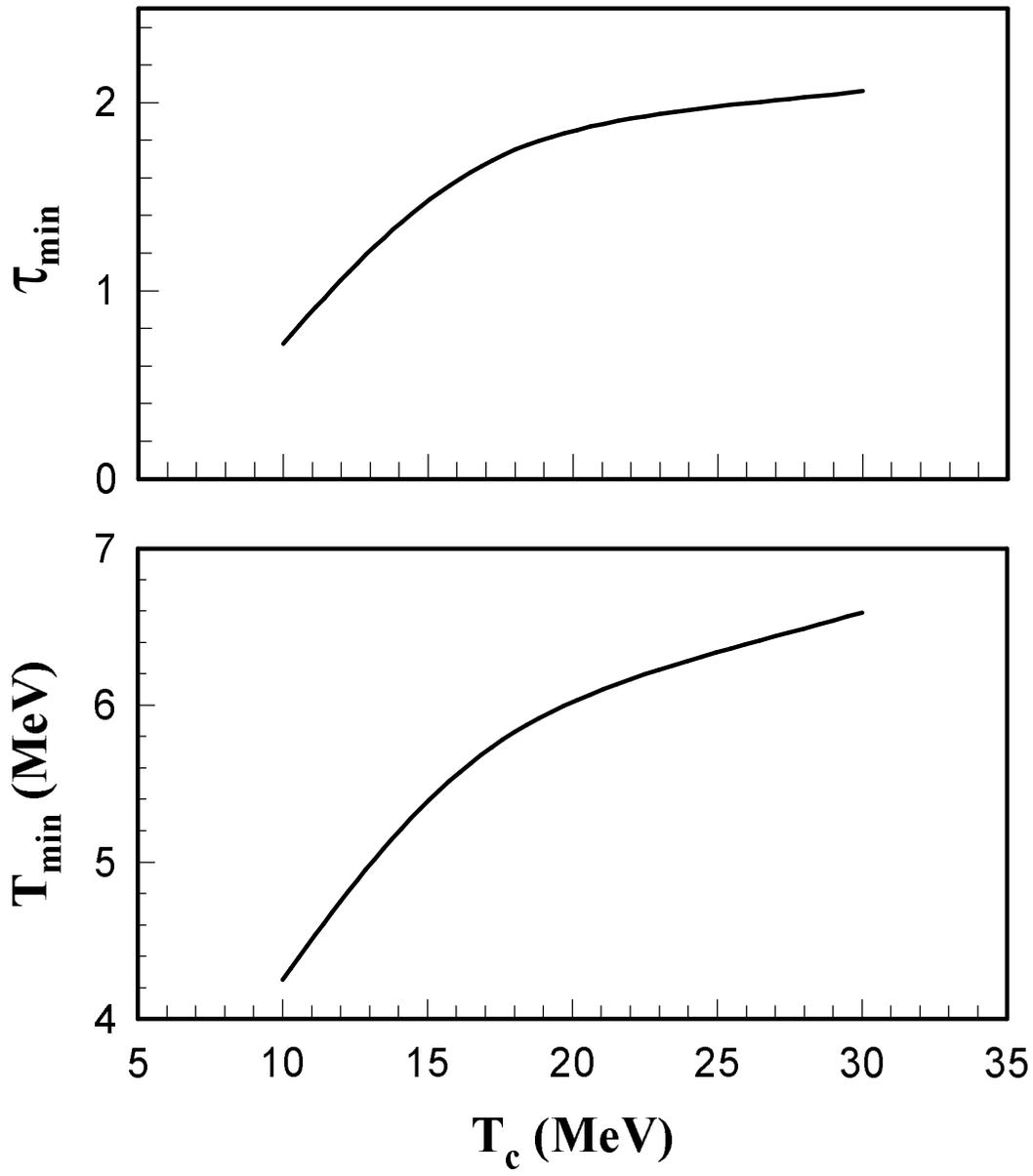
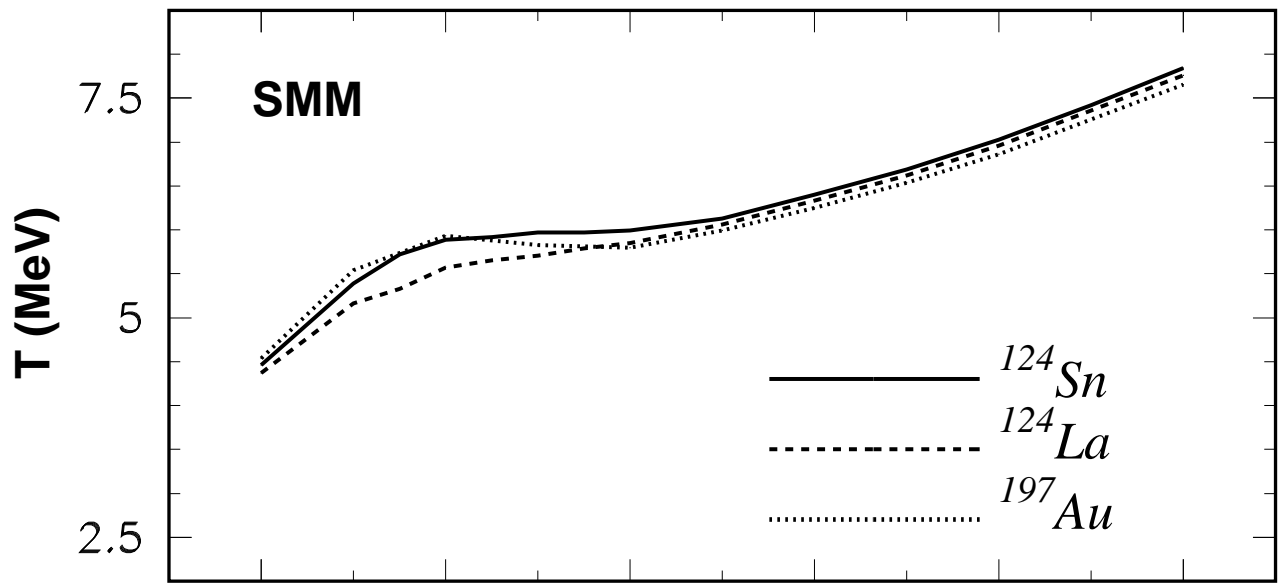
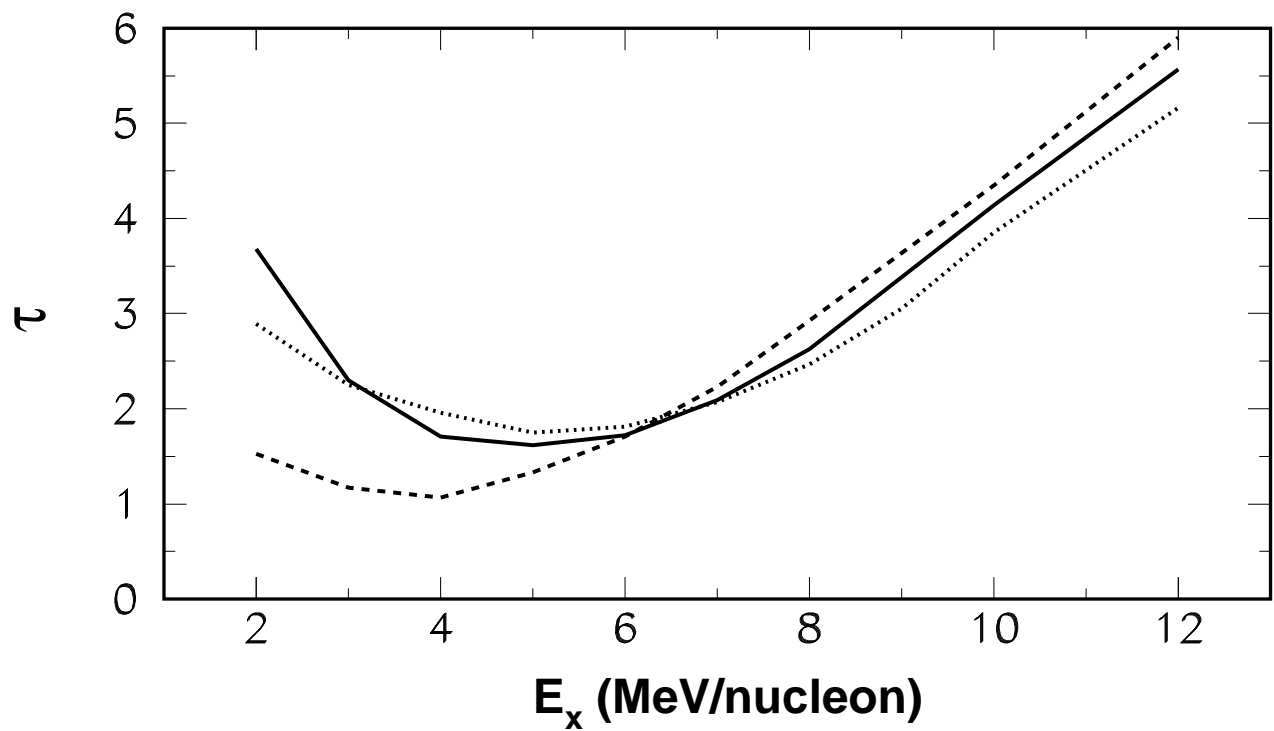
\end{document}